\documentclass[aps,twocolumn,superscriptaddress,nofootinbib]{revtex4-1}
\usepackage{amsmath}
\usepackage{amsfonts}
\usepackage{amssymb}
\usepackage{mathtools}
\usepackage{graphicx}
\usepackage[caption=false,position=t]{subfig}
\usepackage[colorlinks]{hyperref}
\usepackage[capitalize]{cleveref}
\usepackage{color}
\usepackage{multirow}

\usepackage{yypreamble}
\newcommand{\Nq}{N}
\newcommand{\gsz}{{\hat\sigma}^{\rm z}}
\newcommand{\va}{\vec a}
\newcommand{\vb}{\vec b}
\newcommand{\vm}{\vec i}
\newcommand{\dumm}{i}

\newcommand{\vn}{\vec j}
\newcommand{\dumn}{j}

\newcommand{\sig}[1]{{\hat \sigma^{\rm #1}}}

\newcommand{\NU}{N_{\rm U}}
\newcommand{\NM}{N_{\rm M}}
\newcommand{\Ntot}{N_{\rm tot}}
\newcommand{\Dtom}{\Delta_{\rm tom}}

\DeclareMathOperator*{\av}{avg}

\begin{document}

\title{Fourier-style Quantum State Tomography and Purity Measurement of a Multi-qubit System from Bloch Rotations}

\author{Yariv Yanay}
\email{yariv@lps.umd.edu}
\author{Charles Tahan}
\affiliation{Laboratory for Physical Sciences, 8050 Greenmead Dr., College Park, MD 20740}


\begin{abstract}
Quantum state tomography and other measures of the global properties of a quantum state are indispensable tools in understanding many body physics through quantum simulators. Unfortunately, the number of experimental measurements of the system required to estimate these global quantities scales exponentially with system size. Here, we consider the use of random-axis measurements for quantum state tomography and state purity estimation. We perform a general analysis of the statistical deviation in such methods for any given algorithm. We then propose a simple protocol which relies on single-pulse X/Y rotations only. We find that it reduces the basis of the exponential growth, calculating the statistical variance to scale as $\sum_{\va,\vb}\abs{\Delta\rho_{\va\vb}}^{2}\sim 5^{\Nq}/\Ntot$ for full tomography, and $\p{\Delta\mu}^{2}\sim 7^{\Nq}/\Ntot^{2}$ for purity estimation, for $\Nq$ qubits and $\Ntot$ measurements performed.
\end{abstract}

\maketitle

\section{Introduction}

Modern quantum computing platforms have evolved rapidly in the last few years, achieving remarkably long coherence time and preparation and readout fidelity.
While fully robust digital quantum computation is still some way away, near-term devices offer possibilities for new physical insights, including analog quantum simulators that can be used to study many body quantum systems \cite{Bernien2017,Arguello-Luengo2020,Kjaergaard2020,Yanay2020a}.
A quantum simulator, however, is only as good as the information that one can extract from it. While reading out one or a few qubits is sufficient for many purposes, probing strongly interacting systems requires characterizing the many body wavefunction, using tools such quantum state tomography (QST) or the entanglement entropy \cite{Vidal2003,Li2008}.
The number of measurements required to access such quantities scales exponentially with system size. The standard protocol for QST requires $\Ntot \sim 6^{\Nq}$ measurements for an $\Nq$-qubit system \cite{Haah2017}, and the naive approach to evaluating state purity does so via QST and so scales as $\Ntot > 3^{\Nq}$ at least.
A recent proposal \cite{Elben2018,Vermersch2018,Brydges2019,Elben2019} utilized Haar-random unitary rotations to evaluate these quantities, numerically finding a reduced exponential growth rate.

\begin{figure}[tbp] 
   \centering
   \includegraphics[width=\columnwidth]{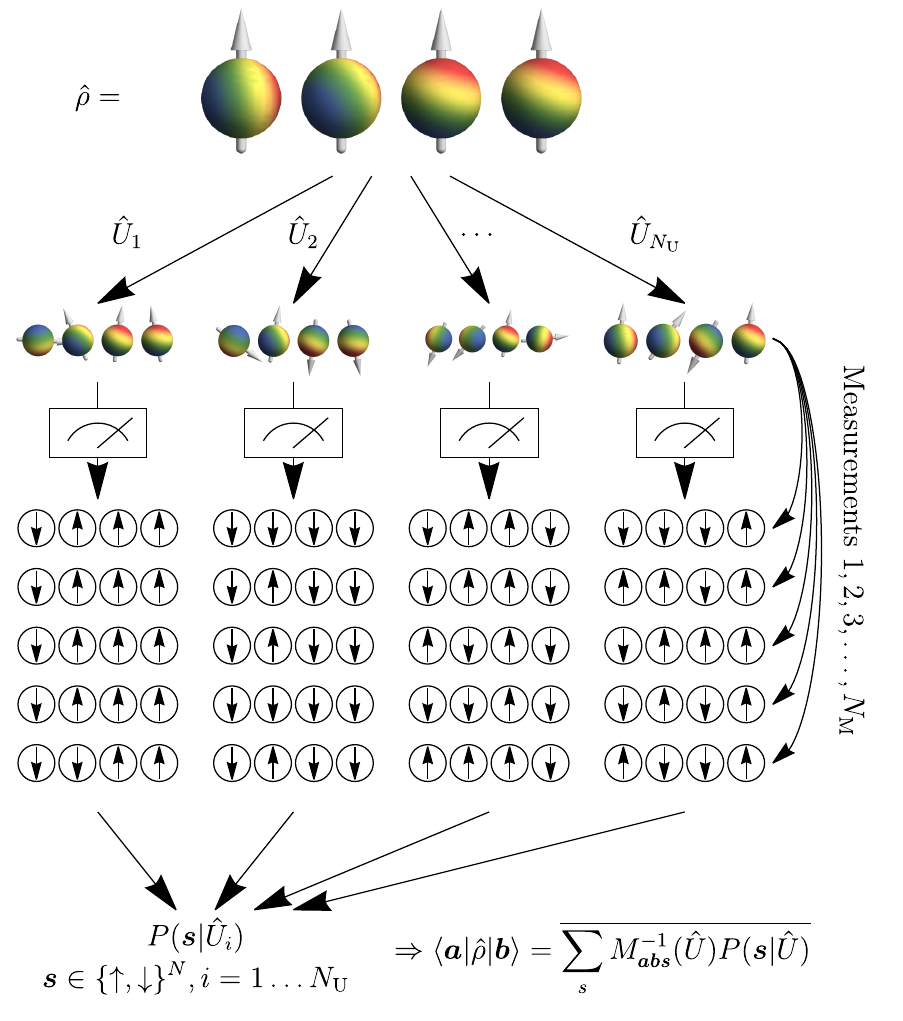} 
   \label{fig:sketch}
   \caption{Quantum state tomography (QST) via random rotations. From top to bottom: an $\Nq$-qubit system is repeatedly prepared in some state $\hat \rho$; we apply each of $\NU$ unitary rotations, randomly sampled from some distribution; for each unitary, we perform $\NM$ measurements of the state of the system, measuring all qubits in the lab Z axis; from all of this we deduce the probability map $P(\vec s \vert \hat U_{i})$ [see \cref{eq:Psabs}]. We then apply an inverse transformation $M^{-1}$ [see \cref{eq:Minvabs}] and average over all rotations to obtain the density matrix.
    }
\end{figure}

Here, we expand on the idea of using random-rotation measurements, formulating a general framework for analyzing random-measurement protocols and their scaling properties, and proposing a protocol tailored for qubits and qubit-like experiments that uses simple rotations along a single X-Y axis. The concept is sketched out in \cref{fig:sketch}.

Making use of the frequency of rotation for each qubit, we extract the state information in Fourier space, performing full state tomography.
We also calculate analytically the statistical variation of the results, finding that the total number of measurements required scales as $\Ntot\sim 5^{\Nq}$, improving over the standard method, and approaching $\Ntot\sim 4^{\Nq}$ one might expect given the amount of information extracted. We also show that our method can be used to measure the state purity, with the total number of measurements scaling as $\Ntot\sim\sqrt{7}^{\Nq} \approx 2.65^{\Nq}$, a significant improvement.

The remainder of the paper is arranged as follows. In \cref{sec:genUs}, we consider the general properties of QST via random-axis measurements. In \cref{sec:XYtom} we propose a simple algorithm using for performing full QST via simple X/Y rotations, and calculate the statistical variation and the number of measurements required. In \cref{sec:mu} we then do the same for measurement of the state purity. Finally, in \cref{sec:limitcon} we consider the case of a control-limited system, where a single drive rotates all qubits.

\section{Characterization via Random Unitaries\label{sec:genUs}}

Consider a system of $\Nq$ qubits which can be repeatedly prepared in a state described by the density matrix
\begin{equation}
\hat \rho_{0} = \sum_{\mathclap{\va,\vb\in \acom{\up,\dn}^{\Nq}}}\rho_{\va\vb}\ket{\va}\bra{\vb}.
\end{equation}
We seek to evaluate the terms $\rho_{\va\vb}$ to some arbitrary accuracy using a minimal number of measurements. We describe a single measurement as the application of a unitary rotation $\hat U$ followed by collapsing the spin state of each qubit along the Z axis and recording the result\footnote{Note that using this terminology, measuring other operators (e.g., $\sig{x}$), is understood as executing a rotation into the appropriate axis (e.g., $\hat U = \exp\br{i\pi\sig{y}/4}$) and then measuring $\sig{z}$.}. 
The total number of measurements, $\Ntot = \NU\times \NM$, consists of the number of different unitaries, $\NU$, multiplied by the number of runs per unitary, $\NM$, required to obtain sufficient statistics for evaluating the probability distribution of measuring each state for a given $\hat U$.

A naive approach can be formulated by observing that each term $\rho_{\va\vb}$ can be obtained from an observable constructed from the product of the operators $1\pm \sig{z}_{q}, \sig{x}_{q}\pm i\sig{y}_{q}$,
\begin{equation}
\rho_{\va\vb} = \avg{\prod_{q=1}^{\Nq}\hat S^{a_{q}b_{q}}_{q}}, \quad 
\begin{gathered}\hat S_{q}^{\up\up}/\hat S_{q}^{\dn\dn} = \frac{1 \pm \hat \gs^{\rm z}_{q}}{2} \\ 
\hat S_{q}^{\dn\up}/\hat S^{\up\dn}_{q} = \frac{\hat \gs^{\rm x}_{q} \pm i\hat \gs^{\rm y}_{q}}{2},\end{gathered}
\label{eq:rhoabnaive}
\end{equation}
where $\sig{x,y,z}_{q}$ are the Pauli operators for qubit $q$. Full state tomography can then be performed by measuring every combination of these Pauli operators, a total of $3^{\Nq}$ different rotations. The number of measurements required for this method can be estimated by calculating the total variance,
\begin{equation}
\Delta_{\rm tom}^{2} \equiv \sum_{\va,\vb}\abs{\Delta \rho_{\va\vb}}^{2},
\end{equation}
which given the number of operators can be estimated as 
\begin{equation}\begin{split}
(\Delta^{\rm naive}_{\rm tom})^{2} 
	& = \frac{1}{\NM}\sum_{\va,\vb}\avg{\prod\abs{\hat S^{\ga_{q}\gb_{q}}_{q}}^{2}} - \abs{\avg{\prod\hat S^{\ga_{q}\gb_{q}}_{q}}}^{2}
	\\ 
	&  = \p{2^{\Nq} - \mu}/\NM.
\end{split}\end{equation}
Here, $\mu=  \sum\abs{\rho_{\va\vb}}^{2} \le 1$ is the state purity, negligible for large $\Nq\gg 1$. We find the total number of measurements scales with system size as 
\begin{equation}
\Ntot^{\rm naive} = \frac{6^{\Nq}}{\Delta_{\rm tom}^{2}}.
\label{eq:NtotNaive}
\end{equation}

\subsection{State Tomography}

To improve on the naive result, we consider the use of randomly sampled measurements. Parameterizing the set of $2^{\Nq}$-dimensional unitaries via some vector of parameters $\vec t$, we sample $\NU$ unitaries with some distribution $\mathcal P_{\vec t}$. As we then measure along the Z axis, the probability of finding the system in a state $\vec x\in\acom{\up,\dn}^{\Nq}$ is given by some map $M$,
\begin{equation}
P_{\vec s}\p{\vec t} = \bra{\vec s}\hat U_{\vec t}\hat\rho_{0} \hat U_{\vec t}\dg\ket{\vec s} \equiv
	\sum_{\mathclap{\va,\vb}}  M_{\vec s \va \vb}\p{\vec t}\rho_{\va\vb}.
\label{eq:Psabs}
\end{equation}

To perform QST, we define an inverse map, $M^{-1}_{\va\vb\vec s}\p{\vec t}$, so that its sample-weighted average over the distribution $\mathcal P_{\vec t}$ inverts the transformation,
\begin{equation}\begin{gathered}
\overline{I_{\va\vb}^{\vm\vn}} = \gd^{\vm}_{\va}\gd^{\vn}_{\vb},
\qquad I_{\va\vb}^{\vm\vn} \equiv {\sum_{\vec s}M^{-1}_{\va\vb\vec s}\p{\vec t}M_{\vec s\vm\vn}\p{\vec t}}, 
\label{eq:Minvabs}
\end{gathered}\end{equation}
where $\overline{f} \equiv \intrm{d\vec t} \mathcal P_{\vec t} f\p{\vec t}$ and  $\gd^{\vm}_{\va}$ is the $\Nq$-dimensional Kronecker delta.

From here, it follows that the set of operators
\begin{equation}
\hat R_{\va\vb}\p{\vec t} = \sum_{\vec x}M^{-1}_{\va\vb\vec x}\p{\vec t}\hat U_{\vec t}\ket{\vec x}\bra{\vec x}\hat U_{\vec t}\dg,
\label{eq:rhoababstract}
\end{equation}
have a sample-weighted expectation value
\begin{equation}
\overline{\avg{\hat R_{\va\vb}}} = \rho_{\va\vb}.
\label{eq:Rababs}
\end{equation}

We note the similarity of this procedure to the concept of shadow tomography \cite{Aaronson2020,Huang2020}. The linear map $M$ projects the $2^{2\Nq}$ terms of the density matrix into the classical shadow containing the $2^{\Nq}$ observable terms. An alternate way of understanding the tomography procedures we describe is as the application of the inverse map to the measurements obtained from multiple random unitaries.

As above, we consider the statistical variance in the result for a given measurements. From \cref{eq:rhoababstract,eq:Rababs}, we find that sampling $\NU$ unitaries and measuring each $\NM$ times, we have
\begin{equation}\begin{split}
\Delta_{\rm tom}^{2} = \frac{1}{\NU}\p{\Delta_{\rm tom}^{\rm U}}^{2} + \frac{1}{\NU\NM}\p{\Delta_{\rm tom}^{\rm M}}^{2} 
\label{eq:dRabAbsdef}
\end{split}\end{equation}
where
\begin{equation}\begin{gathered}
\p{\Delta_{\rm tom}^{\rm U}}^{2} = \sum_{\va,\vb}\br{\overline{\avg{\hat R_{\va\vb}\dg}\avg{\hat R_{\va\vb}}} - \overline{\avg{\hat R_{\va\vb}\dg}}\;\overline{\avg{\hat R_{\va\vb}}}},
\\ \p{\Delta_{\rm tom}^{\rm M}}^{2} = \sum_{\va,\vb}\br{\overline{\avg{\hat R_{\va\vb}\dg\hat R_{\va\vb}}}-\overline{\avg{\hat R_{\va\vb}\dg}}\;\overline{\avg{\hat R_{\va\vb}}}}.
\end{gathered}\end{equation}
It is interesting to note here both terms scale as $1/\NU$, while only the second term changes with $\NM$. As the total number of experimental runs is ${\Ntot = \NU\times\NM}$, the optimal strategy is apparently to maximize the number of unitaries $\NU$ and perform only a small number of measurements $\NM$ of each. Note that $\NM \gg 1$ is required for \cref{eq:dRabAbsdef} to hold, suggesting an ideal $\NM\approx 10$.

\begin{table*}[htbp]
   \centering
   \renewcommand{\arraystretch}{2}
   \setlength{\tabcolsep}{6pt}

   \begin{tabular}{clcrl} 
      	Deviation  & & Form & & Function $W$ 
	\\ \toprule
		$\p{\Delta \mu_{1,0}}^{2}$ 
		& $= 4\Big[\sum\overline{\avg{\hat R_{\vm\vn}}\avg{\hat R_{\vm\pr\vn\pr}}}\rho_{\vn\vm}\rho_{\vn\pr\vm\pr}-\mu^{2}\Big]$
		& \multirow{2}{*}{$\sum W^{\va\vb\va\pr\vb\pr}_{\vm\vn\vm\pr\vn\pr} \rho_{\va\vb}\rho_{\va\pr\vb\pr}\rho_{\vn\vm} \rho_{\vn\pr\vm\pr}$}
		& \multirow{2}{*}{$W^{\va\vb\va\pr\vb\pr}_{\vm\vn\vm\pr\vn\pr}=$}
		& $4\overline{I_{\vm\vn}^{\va\vb}I_{\vm\pr\vn\pr}^{\va\pr\vb\pr}}$
	\\
		$\p{\Delta \mu_{2,0}}^{2}$
		& $=2\Big[\sum \overline{\avg{\hat R_{\vm\vn}}\avg{\hat R_{\vm\pr\vn\pr}}}\;\overline{\avg{\hat R_{\vn\vm}}\avg{\hat R_{\vn\pr\vm\pr}}}-\mu^{2}\Big]$
		& &
		& $2\smashoperator{\sum\limits_{\vec x,\vec y,\vec x\pr,\vec y\pr}}
		\overline{I_{\vec x\vec y}^{\va\vb}I_{\vec x\pr\vec y\pr}^{\va\pr\vb\pr\vphantom{j}}}\; \overline{I_{\vec y\vec x}^{\vn\vm}I_{\vec y\pr\vec x\pr}^{\vn\pr\vm\pr}}$
	\\
		$\p{\Delta \mu_{1,1}}^{2}$ 
		& $= 4\Big[\sum\overline{\avg{\hat R_{\vm\vn}\hat R_{\vm\pr\vn\pr}}}\rho_{\vn\vm}\rho_{\vn\pr\vm\pr}-\mu^{2}\Big]$
		& \multirow{2}{*}{$\sum W^{\va\vb}_{\vm\vn\vm\pr\vn\pr} \rho_{\vb\va}\rho_{\vn\vm} \rho_{\vn\pr\vm\pr}$}
		& \multirow{2}{*}{$W^{\va\vb}_{\vm\vn\vm\pr\vn\pr}=$}
		& $4\overline{J_{\vm\vn\vn\pr\vm\pr}^{\va\vb}}$
	\\
		$\p{\Delta \mu_{2,1}}^{2}$ 
		& $= 4\Big[\sum\overline{\avg{\hat R_{\vm\vn}\hat R_{\vm\pr\vn\pr}}}\;\overline{\avg{\hat R_{\vn\vm}}\avg{\hat R_{\vn\pr\vm\pr}}}-\mu^{2}\Big]$
		& &
		& $4\smashoperator{\sum\limits_{\vec x,\vec y,\vec x\pr,\vec y\pr}}
		\overline{J_{\vec x\vec y\vec x\pr\vec y\pr}^{\va\vb\vphantom{j}}}\; \overline{I_{\vec y\vec x}^{\vn\vm}I_{\vec y\pr\vec x\pr}^{\vn\pr\vm\pr}}$
	\\
		$\p{\Delta \mu_{2,2}}^{2}$ 
		& $= 2\Big[\sum\overline{\avg{\hat R_{\vm\vn}\hat R_{\vm\pr\vn\pr}}}\;\overline{\avg{\hat R_{\vn\vm}\hat R_{\vn\pr\vm\pr}}}-\mu^{2}\Big]$
		& $\sum W^{\va\vb}_{\vm\vn}\rho_{\vb\va}\rho_{\vn\vm} $
		& $W^{\va\vb}_{\vm\vn}= $
		& $2\smashoperator{\sum\limits_{\vec x,\vec y,\vec x\pr,\vec y\pr}}\overline{J_{\vec x\vec y\vec x\pr\vec y\pr}^{\va\vb\vphantom{j}}}\; 
			\overline{J_{\vec y\vec x\vec y\pr\vec x\pr}^{\vn\vm}}$
   \end{tabular}
   \caption{Calculations of bounds and estimates for the variance in the measured purity, as discussed in \cref{sec:purabs}. Wor each deviation term we show the leading contribution, being a weighted sum of quadratic, cubic or quartic combinations of density matrix elements. Wor each type of sum, we show the maximum bound, and the coherent combinations that yield the likely bound and amortized estimate. Summation here is to be taken over all repeated indices.}
   \label{tab:dmuappx}
\end{table*}

The variation with the number of unitaries is
\begin{equation}\begin{gathered}
\p{\Dtom^{\rm U}}^{2} = \sum_{\mathclap{\vm,\vn,\vm\pr,\vn\pr}}\rho_{\vm \vn}\rho_{\vn\pr \vm\pr}W^{\rm U}_{\vm\vn\vn\pr\vm\pr} - \mu, 
\\ W^{\rm U}_{\vm\vn\vn\pr\vm\pr} = \sum_{\vec x,\vec y}\overline{I_{\vec x\vec y}^{\vm\vn}I_{\vec y\vec x}^{\vn\pr\vm\pr}}.
\label{eq:dtomUdef}
\end{gathered}\end{equation}
It is bounded from above by
\begin{equation}\begin{split}
\p{\Dtom^{\rm U}}^{2} & \le \sum_{\mathclap{\vm,\vn,\vm\pr,\vn\pr}}\abs{\rho_{\vm\vn}\rho_{\vn\pr\vm\pr}W^{\rm U}_{\vm\vn\vn\pr\vm\pr} }
	\\ & \le \Big({\sum_{\mathclap{\vm,\vn,\vm\pr,\vn\pr}}\abs{\rho_{\vm\vn}\rho_{\vn\pr\vm\pr}}}\Big)
	\max_{\vm,\vn,\vm\pr,\vn\pr} \abs{W^{\rm U}_{\vm\vn\vn\pr\vm\pr} }.
\label{eq:dtomUmax}
\end{split}\end{equation}

The value of $W^{U}$ in the last line depends on the specific protocol; below, we show that it can be reduced to $\abs{W^{\rm U}_{\vm\vn\vn\pr\vm\pr}}\le 3^{\Nq}$. The sum term, generically, can be as big as $4^{\Nq}$, growing exponentially with system size. However, we note that the sum in \cref{eq:dtomUdef} is composed of terms of essentially random phases arising from the density matrix and $W^{\rm U}$. We might therefore expect it to be well approximated by the coherent terms only,
\begin{equation}\begin{split}
\p{\Dtom^{\rm U}}^{2} & \approx \sum_{\mathclap{\vm,\vn}}\abs{\rho_{\vm \vn}}^{2}W^{\rm U}_{\vm\vn\vn\vm} + \sum_{\vm,\vn}\rho_{\vm \vm}\rho_{\vn\vn}W^{\rm U}_{\vm\vm\vn\vn}
\\ & \le \mu \max_{\vm,\vn}W^{\rm U}_{\vm\vn\vn\vm} + \max_{\vm,\vn}W^{\rm U}_{\vm\vm\vn\vn}.
\label{eq:dtomUest}
\end{split}\end{equation}
Here, $\mu$ is the state purity, which does not grow with with system size. We have therefore reduced the statistical deviation exponentially by the number of qubits.

In fact, amortized over the ensemble of density matrices, we expect the typical variance to be even smaller. To estimate it, we replace the density matrix with its ensemble average, $\abs{\rho_{\vm\vn}}^{2} \to \mu/4^{\Nq}, \rho_{\vm\vm}\to 1/2^{\Nq}$ in \cref{eq:dtomUest}, to find
\begin{equation}\begin{split}
\p{\Dtom^{\rm U}}^{2}_{\rm amo} & \approx \mu \av_{\vm,\vn}{W^{\rm U}_{\vm\vn\vn\vm}} + \av_{\vm,\vn}{W^{\rm U}_{\vm\vm\vn\vn}}.
\label{eq:dtomUamo}
\end{split}\end{equation}
The scaling, in this case, depends only on the function $W^{U}$, and we find in \cref{sec:XYtom} that it is indeed exponentially better than predicted by the upper bound.

A similar calculation can be used to find the variation with the number of measurements per unitary,
\begin{equation}\begin{split}
\p{\Dtom^{\rm M}}^{2} & = 
	\sum_{\mathclap{\vm,\vn}}\rho_{\vm\vn}W^{\rm M}_{\vm\vn} - \mu,
\label{eq:dtomM}
\end{split}\end{equation}
where
\begin{equation}\begin{gathered}
W^{\rm M}_{\vm\vn} = \sum_{\vec x,\vec y}\overline{J_{\vec x\vec y\vec y\vec x}^{\vm\vn}},
\quad J _{\va\vb\vec c\vec d}^{\vm\vn} \equiv \sum_{\vec s}{M_{\vec a\vec b\vec s}^{-1}M_{\vec c\vec d\vec s}^{-1}M_{\vec s\vm\vn}}.
\end{gathered}\end{equation}
We repeat the calculations above to find
\begin{subequations}\label{eq:dtomM}\begin{gather}
\p{\Dtom^{\rm M}}^{2} \le 2^{\Nq}\max_{\vm,\vn}\abs{W^{\rm M}_{\vm\vn}},
\label{eq:dtomMmax}
\\ \p{\Dtom^{\rm M}}^{2} \lesssim \max_{\vm}W^{\rm M}_{\vm\vm},
\label{eq:dtomMest}
\\ \p{\Dtom^{\rm M}}^{2}_{\rm amo} \approx \av_{\vm}W^{\rm M}_{\vm\vm}.
\label{eq:dtomMamo}
\end{gather}\end{subequations}

\subsection{State Purity\label{sec:purabs}}

We consider next a more specialized measurement, applying the technique discussed above to obtaining the state purity,
\begin{equation}
\mu = \Tr\hat\rho^{2} = \sum_{\va,\vb}\rho_{\va\vb}\rho_{\vb\va}.
\end{equation}
The state purity is a measure of the quantum coherence of the state, ranging from $\mu = 1$ for a pure state density matrix $\hat\rho = \ket{\psi}\bra{\psi}$ down to $\mu = 1/2^{\Nq}$ for an infinite temperature mixed state $\hat \rho = \sum_{\vec\ga}1/2^{\Nq}\ket{\gs}\bra{\gs}$. Its logarithm is proportional to the second R\'enyi entropy, and so measuring the purity of subsystems is particularly useful in analyzing global properties of a state such as the entanglement entropy \cite{Eisert2010}.

It follows immediately from \cref{eq:Rababs} that
\begin{equation}
\sum_{\va,\vb}\overline{\avg{\hat R_{\va\vb}}}\;\overline{\avg{\hat R_{\vb\va}}} = \mu.
\label{eq:mumeas}
\end{equation}
The only subtlety in this formulation is that the unitary-averaging for the two terms must be independent. Experimentally, given $\NU$ random unitaries, we sum over all distinct pairs $\vec t_{i},\vec t_{j}$,
\begin{equation}\begin{split}
& \overline{\avg{\hat R_{\va\vb}}}\;\overline{\avg{\hat R_{\vb\va}}} \to
	 \frac{1}{\NU\p{\NU-1}}\sum_{\substack{i,j=1\\ i\ne j}}^{\NU}\avg{\hat R_{\va\vb}\p{\vec t_{i}}}\avg{\hat R_{\vb\va}\p{\vec{t_{j}}}},
\end{split}\end{equation}
omitting the same-unitary $i=j$ term.

We focus, then, on the variance of the measurement. As before, we can calculate the statistical variance, which here includes higher order terms. Taking $\NU,\NM\gg 1$, we find
\begin{equation}\begin{split}
\p{\Delta \mu^{\rm meas}}^{2} = \sum_{m=1}^{2}\sum_{n=0}^{m}\frac{1}{\NU^{m}}\frac{1}{\NM^{n}}\p{\Delta\mu_{{m,n}}}^{2}.
\label{eq:dmudef}
\end{split}\end{equation}
The various deviations $\p{\Delta\mu_{m,n}}^{2}$ are summarized in \cref{tab:dmuappx}.
The value of each of these takes the form of a sum of quadratic, cubic of quartic combinations of density matrix elements, weighted by some function $W$. As in the previous section, the true bound is then given by taking $W\to \max \abs{W}$; and the likely bound and amortized estimate are calculated by taking only the coherent portion of the sum and replacing $W\to \max \abs{W}$ and $W\to \av{W}$, respectively. Below, we calculate these explicitly for our proposed protocol.

\section{Tomography with X-Y Rotations\label{sec:XYtom}}

Using the framework outlined above, we now propose a protocol for QST of a many-qubit system using only X-Y rotations along a single axis. We consider a set of $\Nq$ qubits subject to individual control, each qubit $q$ governed by a Hamiltonian
\begin{equation}
\hat H_{q} = \half \omega_{q}\gsz_{q} + \half \p{\gL_{q}\p{t} \hat\gs^{+}_{q} + \gL_{q}^{*}\p{t}\hat\gs^{-}_{q}}
\end{equation}
where $\gsz_{q},\hat\gs^{+}_{q},\hat \gs^{-}_{q}$ are the qubit's Pauli z, raising and lowering operators, respectively, $\omega_{q}$ its frequency, and $\gL_{q}\p{t}$ some tunable pulse shape.

In applying single-qubit operations, $\gL_{q}$ can be be modulated to produce any kind of rotation. Most simply, however, a single pulse at a constant phase, detuned by $\nu_{q}$ from the qubit frequency, can generate a rotation around a single Bloch vector. In this scenario,
\begin{equation}
\gL_{q}\p{t} = \fopt{e^{-i\p{\omega_{q} + \nu_{q}} t}e^{-i\phi_{q}}g_{q} & t\ge0 \\ 0 & t<0,} 
\label{eq:rot}
\end{equation}
and in the rotating frame at the driving frequency,
\begin{equation}
e^{-i\hat H_{q}t} = e^{-i\gl_{q}t/2}\ket{+}\bra{+}_{q} + e^{i\gl_{q}t/2}\ket{-}\bra{-}_{q},
\end{equation}
with
\begin{equation}\begin{gathered}
\ket{\pm}_{q} = \frac{1}{\sqrt{2}}\br{\sqrt{1 \pm \frac{\nu_{q}}{\gl_{q}}}e^{i\phi_{q} /2}\ket{\dn}_{q}  \pm \sqrt{1\mp \frac{\nu_{q}}{\gl_{q}}}e^{-i\phi_{q} /2}\ket{\up}_{q} },
\\ \gl_{q} = \sqrt{g_{q}^{2} + \nu_{q}^{2}}.
\end{gathered}\end{equation}

We consider a set of measurement at a fixed $g_{q},\nu_{q}$, where the parameters of the transformations $\vec t$ are simply the rotation times $t_{q}$.
The global rotation is given by 
\begin{equation}
\hat U_{\vec t} = \prod_{q}e^{-i\hat H_{q}t_{q}} = \sum_{\mathclap{\vec \ga\in \acom{+,-}^{\Nq}}}e^{-i\sum_{q}\gl_{q}\ga_{q}t_{q}/2} \ket{\vec \ga}\bra{\vec \ga}.
\end{equation}
If we then measure the system's state in the lab basis, the probability of of finding it in a state $\vec s\in\acom{\up,\dn}^{\Nq}$ is given, as in \cref{eq:Psabs}, by $P_{\vec s}\p{\vec t} = \sum_{{\va,\vb}}  M_{\vec s \va \vb}\p{\vec t}\rho_{\va\vb}$ where
\begin{equation}\begin{split}
& M_{\vec s\va\vb}\p{\vec t} = 
\\ &\prod_{q}\br{\sum_{\mathrlap{ \ga,\ga\pr\in \acom{+,-}}}e^{i\gl_{q}\p{\ga-\ga\pr}t_{q}/2} 
	\braket{s_{q}}{\ga}\braket{\ga}{a_{q}}\braket{b_{q}}{\ga\pr}\braket{\ga\pr}{s_{q}}}.
\label{eq:Pseqrhoab}
\end{split}\end{equation}
Here and throughout the text, we use Greek letters $\vec \ga,\vec \gb,\dotsc$ for states in the $\pm$ basis and Latin letters $\va,\vb, \dotsc$ for states in the $\up,\dn$  lab basis. Where the state identifiers ${\ga_{i} = +,-}$, ${a_{i} = \;\up,\dn}$ appear in formulas, they should be taken to mean ${\ga_{i} = 1,-1}$, ${a_{i} = 1,-1}$, respectively.

Continuing to follow the terminology of \cref{sec:genUs}, we define a set of observable $\hat R_{\va\vb}$ as in \cref{eq:rhoababstract}, by way of the inverse transformation
\begin{equation}\begin{split}
& M^{-1}_{\va\vb\vec s}\p{\vec t} = 
	\prod_{q}e^{i\phi_{q}\p{b_{q}-a_{q}}/2}\Big[\gd_{2s_{q}}^{a_{q}+b_{q}}-\frac{s_{q}}{2}\Big(\frac{g_{q}}{\nu_{q}}\gd^{-b_{q}}_{a_{q}}
		\\ & - e^{i\gl_{q}t_{q}}b_{q}\sqrt{\frac{\gl_{q}+b_{q}\nu_{q}}{\gl_{q}+a_{q}\nu_{q}}} 
		- e^{-i\gl_{q}t_{q}}a_{q}\sqrt{\frac{\gl_{q}+a_{q}\nu_{q}}{\gl_{q}+b_{q}\nu_{q}}}\Big)\Big].
\label{eq:MinvXY}
\end{split}\end{equation}
The transformation is constructed so that it removes all terms that rotate at the same frequency as $\rho_{\va\vb}$,
\begin{equation}\begin{split}
& I_{\va\vb}^{\vm\vn}\p{\vec t} = \sum_{\vec x}M^{-1}_{\va\vb\vec x}\p{\vec t}M_{\vec x\vm\vn}\p{\vec t} = 
\\ & \prod_{q}\br{\gd^{\dumm_{q}}_{a_{q}}\gd^{\dumn_{q}}_{a_{q}} + O\p{e^{\pm i \gl_{q}t_{q}}} + O\p{e^{\pm 2i \gl_{q}t_{q}}}}.
\end{split}\end{equation}
The unwanted terms in the second line of can now be removed by averaging over different rotations. If we uniformly sample from $\vec t\in \br{0,T}^{\Nq}$, we find
\begin{equation}\begin{gathered}
\overline{I_{\va\vb}^{\vm\vn}} = \gd_{\va}^{\vm}\gd_{\vb}^{\vn} + 
	 O\Big( \sum_{q}\frac{1}{\gl_{q} \p{a_{q} - b_{q} - \dumm_{q} + \dumn_{q}}T}\Big).
\label{eq:avRT}
\end{gathered}\end{equation}
For long sampling periods, $\forall q, \gl_{q}T\gg \Nq$, the value of the observable converges to that of the respective term of the density matrix as above,
\begin{equation}
\lim_{T\to \infty}\overline{I_{\va\vb}^{\vm\vn}} = \gd_{\va}^{\vm}\gd_{\vb}^{\vn},
\qquad \lim_{T\to \infty}\overline{\avg{\hat R_{\va\vb}}} = \rho_{\va\vb}.
\label{eq:Rabmeas}
\end{equation}

Following \crefrange{eq:dtomUdef}{eq:dtomM}, we find that the variance scaling with number of unitaries behaves as
\begin{subequations}\begin{gather}
\begin{split} & \p{\Dtom^{\rm U}}^{2} \le \\ & \quad 4^{\Nq}\prod_{q}\max\br{3+S_{q}\frac{g_{q}^{2}}{\gl_{q}^{2}}, 2+S_{q}({1 + \frac{\nu_{q}^{2}}{\gl_{q}^{2}}})},\end{split}
\\ \begin{split}& \p{\Dtom^{\rm U}}^{2}  \lesssim \prod_{q}\br{2+S_{q}({1 + \frac{\nu_{q}^{2}}{\gl_{q}^{2}}})}+
	\\ &\quad  \p{\prod_{q}{\max\br{3+S_{q}\frac{g_{q}^{2}}{\gl_{q}^{2}}, 2+S_{q}({1 + \frac{\nu_{q}^{2}}{\gl_{q}^{2}}})}}}\mu,\end{split}
\\ \p{\Dtom^{\rm U}}^{2}_{\rm amo} \approx \p{\prod_{q}\br{\frac{5}{2} + S_{q}}}\mu + \frac{1}{2^{\Nq}},
\label{eq:dTomUamo}
\end{gather}\end{subequations}
where $S_{q} = \frac{\p{g_{q}^{2}-2\nu_{q}^{2}}^{2}}{4g_{q}^{2}\nu_{q}^{2}}$.
The variance scaling with the total number of measurements is independent of the density matrix, and simply has the form
\begin{equation}
\p{\Dtom^{\rm M}}^{2} =  \prod_{q}\br{5 + 2S_{q}}.
\label{eq:dTomM}
\end{equation}
We observe that there is a sweet spot at $g_{q} = \sqrt{2}\nu_{q}$, $S_{q} = 0$, where the variance is minimized.

From \cref{eq:dRabAbsdef}, we find an estimate for the number of experimental runs required to perform QST using this technique. 
Noting that the controlling variance comes from $\p{\Dtom^{\rm M}}^{2}$, we find at the sweet spot
\begin{equation}
\Ntot^{\rm Fourier} = \frac{5^{\Nq}}{\Delta_{\rm tom}^{2}}.
\label{eq:NtotFourier}
\end{equation}
This is an improvement on the naive estimate of \cref{eq:NtotNaive}, by a factor of $\p{5/6}^{\Nq}$. While this may seem marginal, like all exponential factors it can be quite significant, reducing measurement number by half for as few as four qubits, and by 80\% for nine. The number of measurements required remains exponential, but this is not entirely unexpected; full tomography involves measuring $4^{\Nq}$ terms of the density matrix, and on this ground alone we might expect $\Ntot \geq {4^{\Nq}}/{\Delta_{\rm tom}^{2}}$.

\begin{figure}[tbp] 
   \centering
   \includegraphics[width=\columnwidth]{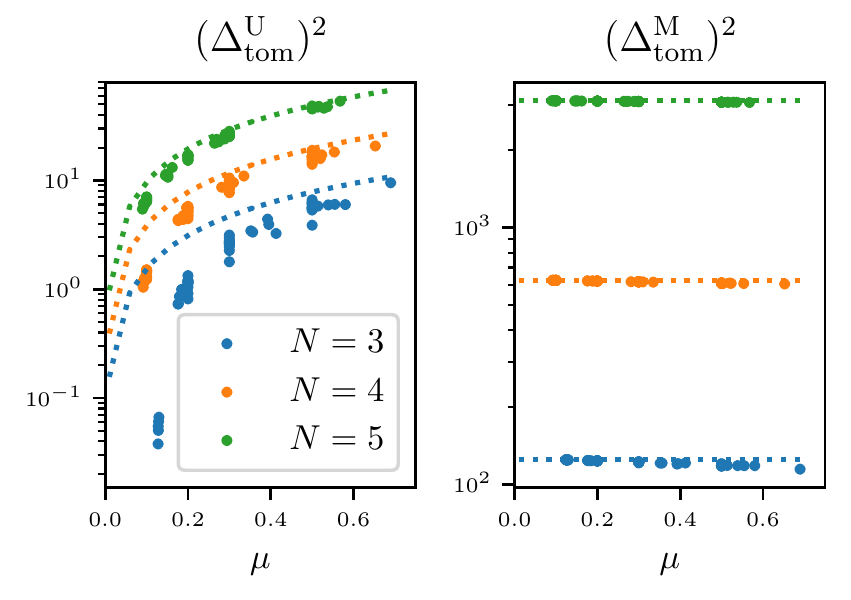} 
   \caption{Scaling of the measurement variance in full QST, including statistical variance depending on the number of unitaries (left) and number of measurements (right), see \cref{eq:dRabAbsdef}. They are shown for 3, 4, and 5 qubits, at the optimal point $g_{q} = \sqrt{2}\nu_{q}$, for 60 different density matrices each. We generate these density matrices in three different ways, as described in \cref{sec:numerics}. The dotted lines show the amortized estimates given by \cref{eq:dTomUamo}, $\p{\Dtom^{\rm U}}^{2}_{\rm amo} \approx \p{5/2}^{\Nq}\mu$ on the left; and the calculated value of \cref{eq:dTomM}, $\p{\Dtom^{\rm M}}^{2} =  5^{\Nq}$.
   }
   \label{fig:dTom}
\end{figure}

We compare the expected scaling of these deviations to numerical calculations for randomly generated density matrices of different kinds in \cref{fig:dTom}. In all cases we observe that the predicted approximate bounds describe the system well. Our numerical results span the range $0.08\le \mu \le 0.69$, with random density matrices generated in three different ways, and so we expect them to be quite robust. However, it is possible that some subset of states, perhaps correlated with the $\pm$ measurement basis, would exhibit worse scaling.

\section{Purity with X-Y Rotations\label{sec:mu}}

\begin{figure}[tbp] 
   \centering
   \includegraphics[width=\columnwidth]{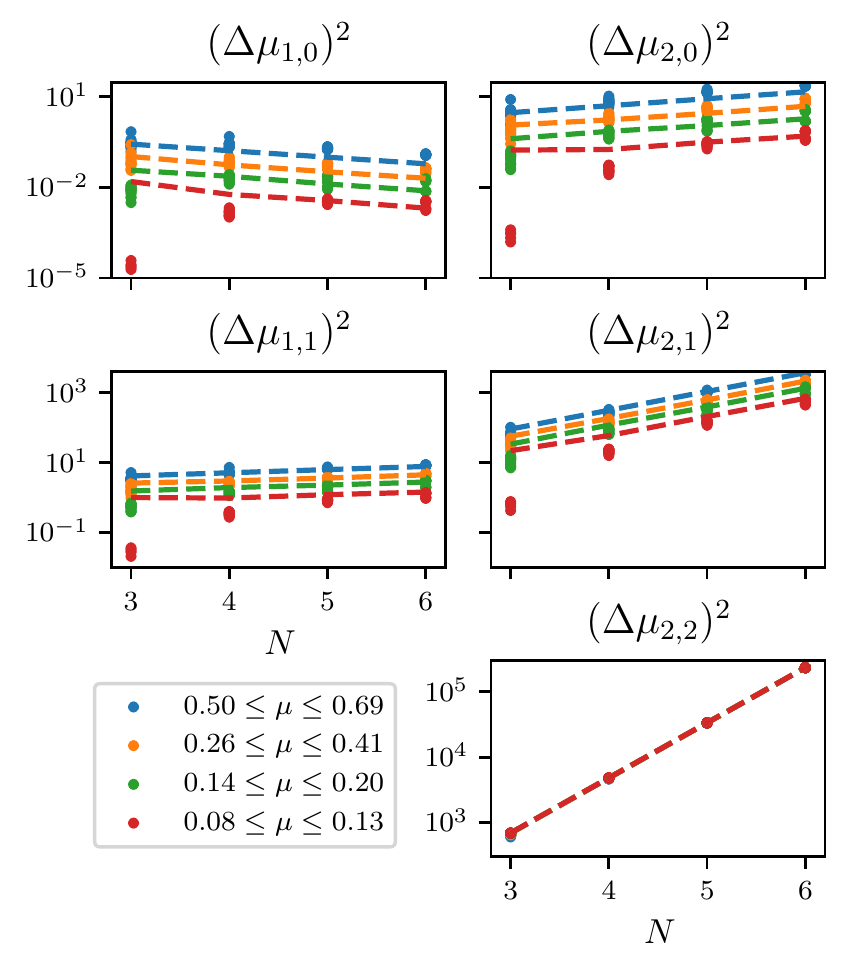} 
   \caption{Scaling of the statistical variance in purity measurement with the number of qubits in the system [see \cref{eq:dmudef}], comparing numerical results and analytical estimates at the optimal working point, $g_{q} = \sqrt{2}\nu_{q}$ for all qubits.
   In each plot, we show the results from each of 60 different density matrices randomly generated in three different ways, as described in \cref{sec:numerics}.
   The dashed lines correspond to the amortized estimates given in \cref{tab:dmuest} at the optimal working point.
   We observe good agreement between our predictions and the numerical calculations.
	}
   \label{fig:dmuAll}
\end{figure}

Next, we apply the same technique to obtaining the state purity. We perform the same calculations as in the previous section to find upper bounds, approximate bounds and amortized estimates for these deviations. We show the leading term for the amortized estimates in \cref{tab:dmuest}. We find that the same sweet spot, at $g_{q}=\sqrt{2}\nu_{q}$, leads to ideal scaling behavior. We compare these estimates to numerical results for a variety of random density matrices in \cref{fig:dmuAll}, finding remarkable agreement with the amortized estimates.

We find, as shown in \cref{tab:dmuest}, that the dominant scaling behavior comes from $\Delta\mu_{2,2}$, which grows as ${\Delta\mu_{2,2} \sim \prod_{q}\br{7+O\p{S_{q}}}^{\Nq}}$. With optimal parameters, $g_{q} = \sqrt{2}\nu_{q}$, we thus expect the total number of experimental runs required to measure the purity to scale as $N_{\rm total} = \NU\times\NM\sim \sqrt{7}^{\Nq}\approx 2.65^{\Nq}$. This is a large improvement over the scaling of measuring the purity via full QST, which requires at least $N_{\rm total}> 3^{\Nq}$ runs without taking into account scaling of its variance. Interestingly, $\sqrt{7}\approx 2^{1.4}$, making this scaling comparable with the results seen in Ref.~\onlinecite{Brydges2019} for non-pure states, such as we used here.

\begin{table}[tbp]
   \centering

   \begin{tabular}{c l c} 
      	Deviation &  Leading term & Optimal Scaling
	\\
      \toprule
	$\p{\Delta \mu_{1,0}}^{2}_{\rm amo}$ & $ = 4\p{\prod_{q}\br{\frac{5}{8}+O\p{S_{q}}}}\mu^{2}$ & $\NU\sim0.63^{\Nq}$
	\\
	$\p{\Delta \mu_{2,0}}^{2}_{\rm amo}$ & $ = 2\p{\prod_{q}\br{\frac{7}{4}+O\p{S_{q}}}}\mu^{2}$ & $\NU\sim1.32^{\Nq}$
	\\
	$\p{\Delta \mu_{1,1}}^{2}_{\rm amo}$ & $= 4\p{\prod_{q} \br{\frac{5}{4}+O\p{S_{q}}}}\mu$ & $\NU\NM\sim1.25^{\Nq}$
	\\
	$\p{\Delta \mu_{2,1}}^{2}$ & $ =4\p{\prod_{q} \br{\frac{7}{2}+ O\p{S_{q}}}}\mu$ & $\NU\NM^{1/2}\sim1.87^{\Nq}$
	\\
	$\p{\Delta \mu_{2,2}}^{2}$ & $ = 2\prod_{q} \br{7+ O\p{S_{q}}}$ & $\NU\NM\sim 2.65^{\Nq}$
   \end{tabular}

   \caption{Estimates of the variances in the measured quantum state purity, as discussed in \cref{sec:mu}. For each deviation term we show the weight function for its leading contribution (see \cref{tab:dmuappx}), and the amortized estimate of its value, as a function of the parameters $g_{q},\nu_{q}$ for each qubit. We also show the dominant scaling term for each row in the optimal case $g_{q} = \sqrt{2}\nu_{q}$. Here $\gd^{\vec x}_{\vec y}$ is the $\Nq$-dimensional Kronecker delta function for all components of $x_{i}-y_{i}$.
   }
   \label{tab:dmuest}
\end{table}

\section{Limited Control Systems\label{sec:limitcon}}

We now turn to consider the application of a model in a system where qubit control is limited. We consider a system inspired by Ref.~\onlinecite{Yanay2020a}, where multiple resonant qubits are coupled to a single signal line via different-frequency resonators, so that individual dispersive readout is available but driving at the qubit frequency affects all qubits concurrently. 

In terms of our model,we take
\begin{equation}
\omega_{q} \to \omega_{0}, \qquad \gL_{q}\p{t} \to \chi_{q}\gL\p{t}.
\end{equation}
In this case, the rotation is applied to all qubits simultaneously and the unitaries are characterized by a single time, $t_{q} \to t$. The analysis of \cref{sec:genUs} generally applies. However, the task of picking out Fourier component becomes slightly more subtle. Returning to \cref{eq:avRT}, we now have just a single integration parameter, and so
\begin{equation}\begin{gathered}
\overline{I_{\va\vb}^{\vm\vn}} = \gd_{\va}^{\vm}\gd_{\vb}^{\vn} + 
	 O\Big( \frac{1}{\theta_{\va-\vb-\vm+\vn}T}\Big).
\label{eq:IsingleT}
\end{gathered}\end{equation}
where
\begin{equation}
\theta_{\va-\vb-\vm+\vn} = \sum_{q}\gl_{q}\p{a_{q}-b_{q} - \dumm_{q} +\dumn_{q}}/2.
\label{eq:lincomb}
\end{equation}

Instead of picking out the desired frequency for each component individually, we are able to select only the overall frequency for the entire system. We are still able to pick a precise component as long as the set of frequencies $\gl_{q}$ in linearly independent under combinations
\begin{equation}\begin{gathered}
\forall \vec x \in \acom{-2,-1,0,1,2}^{\Nq}
\qquad \vec \gl \cdot \vec x = 0 \Leftrightarrow \vec x = \vec 0
\\ \Rightarrow \forall \va,\vb,\vm,\vn, \quad \theta_{\va-\vb-\vm+\vn} = 0 \Leftrightarrow \va-\vb=\vm-\vn.
\end{gathered}\end{equation}
Then, for a sufficiently large $T$, the value of the observable converges again, as in \cref{eq:Rabmeas},
\begin{equation}
\lim_{T\to \infty}\overline{I_{\va\vb}^{\vm\vn}} = \gd_{\va}^{\vm}\gd_{\vb}^{\vn},
\qquad \lim_{T\to \infty}\overline{\avg{\hat R_{\va\vb}}} = \rho_{\va\vb}.
\end{equation}

\begin{figure}[tbp] 
   \centering
   \includegraphics[width=\columnwidth]{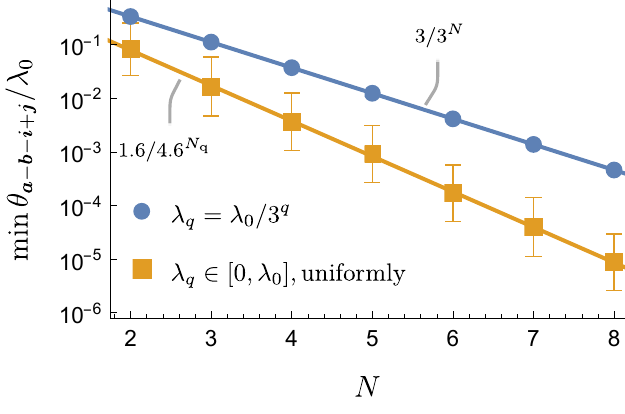} 
   \caption{The smallest linear combination of frequencies, $\theta_{\va-\vb-\vm+\vn}$, as defined in \cref{eq:lincomb}, for an exponential frequency ladder $\gl_{q} = \gl_{0}/3^{q}, q=0,\dotsc,\Nq-1$ (blue circles) and for frequencies randomly chosen from a uniform distribution $\gl_{q}\in\br{0,\gl_{0}}$ (yellow squares). For random frequencies, we average over 200 randomizations and show the logarithmic mean and standard deviation.
   The smallest linear combination decreases with qubit number as $\min \br{\theta} \sim 1/3^{\Nq},1/4.6^{\Nq}$, respectively, implying that the maximum rotation time $T$ must increase exponentially with qubit number. 
   We find that variations on these arrangements such as a uniform offset for all frequencies or non-uniform sampling of frequencies (not shown here) do not significantly change the scaling behavior.}
   \label{fig:mind}
\end{figure}

We can define the systematic deviation of the measurement from the ideal value,
\begin{equation}
\overline{\avg{\hat R_{\va\vb}}} \equiv \rho_{\va,\vb} + \frac{1}{T}\Delta \rho_{\va,\vb}^{\rm sys}.
\end{equation}
From \cref{eq:IsingleT}, it is clear that this deviation scales as
\begin{equation}
\abs{\Delta \rho_{\va,\vb}^{\rm sys}} \sim 1/ \min_{\mathclap{\vm-\vn\ne \va-\vb}}\abs{\theta_{\va-\vb-\vm+\vn}},
\end{equation}
inversely proportional to the smallest linear frequency combination of \cref{eq:lincomb}. This combination varies with the distribution of frequencies, but for a fixed frequency range, $\gl_{i}\in\br{0,\gL}$, we expect it to grow exponentially with the number of qubits $\Nq$. This is borne out in numerical calculations, both for ideally chosen and randomly picked frequencies, as shown in \cref{fig:mind}. A similar calculation applies for the purity $\mu$.

Experimentally, the maximum rotation time $T$ is generally limited by the coherence time of the qubits. 
To measure the terms of of the density matrix, we require $\abs{\Delta\rho_{\va,\vb}^{\rm T}/T} \lesssim \abs{\rho_{\va,\vb}}$. Taking ${\abs{\rho_{\va,\vb}}\approx 1/2^{\Nq}}$, we find
\begin{equation}
T\gtrsim 2^{\Nq}/\min \theta_{\va-\vb-\vm+\vn} \gtrsim 6^{\Nq}/\gl_{0},
\end{equation}
where $\gl_{0}$ the frequency spread of the qubits in the system. 
If it is on the order of $\gl_{0}/2\pi \approx 1\unit{GHz}$, a coherence time of $T\approx 10\unit{\mu s}$ would allow the probing of up to $\Nq = 5-6$ qubits. To measure state purity, a similar calculation shows
\begin{equation}
T\gtrsim 1/\min \theta_{\va-\vb-\vm+\vn} \gtrsim 3^{\Nq}/\gl_{0},
\end{equation}
allowing up to $\Nq = 7-10$  depending on the distribution of frequencies.

\section{Discussion}

We have analyzed here the use of random-axis measurements for characterization of two global characteristics of the wavefunction, quantum sate tomography and estimation of the state purity and obtained upper bounds and average estimates of the scaling behavior with system size. We've presented a protocol for these measurements using simple X/Y rotations, alone, finding an exponential law with a basis that is significantly reduced compared with the naive methods.

We observe that the scaling law we find for measurement of the purity, $\Ntot \sim \sqrt{7}^{\Nq}$ is similar to the result obtained numerically using Haar-random rotations for mixed states similar to the kind we explore here \cite{Brydges2019}. This similarity may hint at some universal scaling properties such random-sampling protocols that could be explored in future work.

Finally, in addition to the reduction in number of measurements required, we note the protocol we suggest here has several experimental advantages. Generating a simple rotation of the form in \cref{eq:rot} is experimentally straightforward, as is the characterization of the frequency offsets $\nu_{q}$ and coupling parameters $g_{q}$. Significantly, the phase parameters in the inverse transformation of \cref{eq:MinvXY} scale linearly with the rotation time, and are therefore not sensitive to the shapes of the rise and termination of the driving pulse as more complicated schemes may be. This protocol could be immediately realized in many existing experimental systems.

\appendix

\section{Generating random density matrices\label{sec:numerics}}

For a specified $\mu$ and qubit number $\Nq$, we generate random density matrices with the desired purity and dimensionality in three different ways, as given below. We perform all calculations using the QuTiP framework \cite{Johansson2013}. For each $\Nq = 3,4,5,6$, $\mu=0.5,0.3,0.2,0.1$ and each of the three generations methods, we generate five different density matrices, for a total of 60 matrices at each $\Nq$.

\subsection{Geometrically weighted Haar-random vectors}
\begin{enumerate}
\item We generate $n = \min\br{2^{\Nq},30}$ states, sampled using the Haar measure over the $2^{\Nq}$-dimensional Hilbert space.
\item We iteratively orthonormalize the set of vectors to generate $\ket{\psi_{m}}$.
\item With $w_{m} = A x^{m}$, and solve ${\sum w_{m} = 1}$, ${\sum w_{m}^{2} = \mu}$ for  $A,x$.
\item We assign
\begin{equation}
\hat\rho = \sum_{m}w_{m}\ket{\psi_{m}}\bra{\psi_{m}}.
\end{equation}
\end{enumerate}

\subsection{Uniformly weighted Haar-random vectors}
\begin{enumerate}
\item We generate $n = \lceil1/\mu\rceil$ (i.e.~$1/\mu$ rounded up) states, sampled using the Haar measure over the $2^{\Nq}$-dimensional Hilbert space.
\item We iteratively orthonormalize the set of vectors to generate $\ket{\psi_{m}}$.
\item We solve $w_{1}^{2} + \p{n-1} \p{\frac{1-w_{1}}{n-1}}^{2} =\mu$ for $w_{1}$.
\item We assign
\begin{equation}
\hat\rho = w_{1}\ket{\psi_{1}}\bra{\psi_{1}} + \frac{1-w_{1}}{n-1}\sum_{m=2}^{n}\ket{\psi_{m}}\bra{\psi_{m}}.
\end{equation}
\end{enumerate}

\subsection{Partial trace from higher-dimensional vector}
\begin{enumerate}
\item With $\overline{\mu_{n}} = 2^{-\Nq} + 2^{-n} - 2^{-n-\Nq}$, find the integer $n\ge 1$ which minimizes $\abs{\mu -\overline{\mu_{n}}}^{2}$.
\item We generate a state $\ket{\psi}$,sampled from a Haar distribution over the $2^{\Nq+n}$-dimensional Hilbert space.
\item We trace out $n$ qubits, assigning
\begin{equation}
\hat \rho = \Tr_{q=1\dotsc n}\ket{\psi}\bra{\psi}.
\end{equation}

\end{enumerate}

\bibliography{FourierTom.bbl}

\end{document}